Michael Deininger, Maximilian Hansen, Jens Fohlmeister, Andrea Schröder-Ritzrau, Yuval Burstyn, Denis Scholz, 2021, Are oxygen isotope fractionation factors between calcite and water derived from speleothems systematically biased due to prior calcite precipitation (PCP)?, Geochimica et Cosmochemica Acta 305, 212-227.

# Are oxygen isotope fractionation factors between calcite and water derived from speleothems systematically biased due to prior calcite precipitation (PCP)?

Michael Deininger[1,*], Maximilian Hansen[1], Jens Fohlmeister[2,3], Andrea Schröder-Ritzrau[4], Yuval Burstyn[5], Denis Scholz[1]

*Corresponding Author: michael.deininger@uni-mainz.de

1) Institute for Geosciences, Johannes Gutenberg University Mainz, Johann-Joachim-Becher-Weg 21, 55128 Mainz, Germany

2) Potsdam Institute for Climate Impact Research, Telegrafenberg, 14473 Potsdam, Germany

3) GFZ German Research Centre for Geosciences, Section 'Climate Dynamics and Landscape Development', Telegrafenberg, 14473 Potsdam, Germany

4) Institute of Environmental Physics, Heidelberg University, Im Neuenheimer Feld 229, 69120 Heidelberg, Germany

5) The Fredy & Nadine Herrmann Institute of Earth Sciences, The Hebrew University, Jerusalem, 9190401, Israel

* Corresponding author

***Abstract:***

*The equilibrium oxygen isotope fractionation factor between calcite and water ($^{18}\alpha_{calcite/H2O}$) is an important quantity in stable isotope geochemistry and allows in principle to infer temperature variations from carbonate $\delta^{18}O$ if carbonate formation occurred in thermodynamic equilibrium. For this reason, many studies intended to determine the value of the oxygen isotope fractionation factor between calcite and water ($^{18}\alpha_{calcite/H2O}$) for a wide range of temperatures using modern cave calcite and the corresponding cave drip water or ancient speleothem carbonate and fluid inclusion samples. However, the picture that emerges from all of these studies indicates that speleothem calcite is not formed in thermodynamic equilibrium but under kinetic conditions, provoking a large variability of determined $^{18}\alpha_{calcite/H_2O}$ values. Here we present a conceptual framework that can explain the variability of $^{18}\alpha_{calcite/H_2O}$ values obtained by cave studies. Prior calcite precipitation (PCP) is calcite precipitation before cave drip water is dripping from the cave ceiling and impinges on the surface of a stalagmite or watch glass. Prior to the karst water dripping from the cave ceiling, PCP can occur in the karst above the cave as well as on the cave ceiling, the cave walls and on the surface of stalactites. We argue that PCP leads to increasing the $\delta^{18}O$ value of the dissolved $HCO_3^-$ ($\delta^{18}O_{HCO3-}$), resulting in an oxygen isotope disequilibrium of the $\delta^{18}O_{HCO3-}$ values with respect to the $\delta^{18}O$ value of water ($\delta^{18}O_{H2O}$). The oxygen isotope disequilibrium between $HCO_3^-$ and $H_2O$ is re-equilibrated by oxygen isotope exchange between $H_2O$ and $HCO_3$. Depending on the temperature, the re-equilibration time varies from hours to days and is usually much longer than the residence time of the drip water on stalactites, but much shorter than the time required to percolate through the karst. Therefore, while the oxygen isotope equilibrium between $HCO_3^-$ and $H_2O$ is very likely re-established when PCP occurred in the karst, oxygen isotope disequilibrium conditions between $HCO_3^-$ and $H_2O$ still prevail when PCP occurred inside a cave, e.g., on stalactites. If the oxygen isotope disequilibrium conditions* *between $HCO_3^-$ and $H_2O$ is not re-established, the precipitated calcite will inherit the elevated $\delta^{18}O$ value of the $HCO_3^-$ and not be in oxygen isotope equilibrium with the corresponding drip water. Consequently, if the $^{18}\alpha_{calcite/H_2O}$ value is calculated from cave calcite samples affected by PCP, the derived value will be systematically biased.*

## 1. Introduction

Coplen (2007) set off a debate questioning the temperature relationship of the oxygen isotope equilibrium fractionation factor between calcite and $H_2O$ ($^{18}\alpha_{calcite/H_2O}$) derived by Kim and O'Neil (1997) (Daëron et al., 2019; Day and Henderson, 2011; Dietzel et al., 2009; Feng et al., 2012a; Feng et al., 2014; Johnston et al., 2013; McDermott et al., 2011; Tremaine et al., 2011). Coplen (2007) observed that the oxygen isotope fractionation for the sub-aqueous carbonate sample from Devils Hole at the given groundwater temperature of 33.7 °C is underestimated by the experimentally derived $^{18}\alpha_{calcite/H2O}$ relationship of Kim and O'Neil (1997) (Coplen, 2007). Since then, a lively debate developed about which oxygen isotope fractionation factor between calcite and $H_2O$ is most accurate. Recent speleothem-based studies support Coplen's (2007) observation (e.g., Feng et al., 2012b; Feng et al., 2014; Johnston et al., 2013; McDermott et al., 2011; Riechelmann et al., 2013; Tremaine et al., 2011) demonstrating that the majority of apparent oxygen isotope fractionation factors calculated from the $\delta^{18}O$ values of speleothem calcite and the corresponding drip water are underestimated by Kim and O'Neil's (1997) $^{18}\alpha_{calcite/H_2O}$ temperature-relationship (Fig. 1). Furthermore, not only cave carbonates but also empirical oxygen isotope fractionation factors obtained from inorganic $CaCO_3$ precipitation experiments appear to support a larger oxygen isotope fractionation factor than the $^{18}\alpha_{calcite/H2O}$ value determined by Kim and O'Neil (1997) (Day and Henderson, 2011; Dietzel et al., 2009; Hansen et al., 2019). Moreover, the laboratory experiments performed by Dietzel et al. (2009) demonstrate that the oxygen isotope fractionation between calcite and water depends, among other parameters, on the pH value of the solution and on the $CaCO_3$ precipitation rate, which has recently also been found for inorganic calcite precipitation experiments from thin films (Hansen et al., 2019).

<<< Figure 1 >>>

In addition to the experimental studies, several theoretical models have been developed to explain the observed variations in oxygen isotope fractionation between calcite and water. DePaolo (2011) described a ‘surface reaction kinetic model’ that considers isotope and trace element fractionation processes during calcite precipitation. The results show that oxygen isotope fractionation is a function of the ratio between calcite precipitation and dissolution rates. Assuming a constant calcite dissolution rate, the oxygen isotope fractionation should represent equilibrium conditions for low net precipitation rates, whereas it should correspond to kinetic conditions in case of high net calcite precipitation rates. This model has been used to explain the relationship of $\delta^{18}O$ values in inorganically precipitated calcite with temperature, pH and growth rate (Watkins et al., 2014). Another concept has been presented by Watson (2004), who described a surface entrapment model based on solid-state diffusion. However, the surface entrapment model may not be relevant for oxygen isotopes in speleothems because the solid-state diffusion rate is too small compared to the calcite precipitation rates (DePaolo, 2011; Hendy, 1971). A further mechanism proposed by Zeebe (1999) and Zeebe (2007) is that the $\delta^{18}O$ value of calcite reflects the mean $\delta^{18}O$ value of the dissolved inorganic carbon (DIC), i.e., of $CO2^*_{aq}$ (incl. $H_2CO_3$), $HCO_3^-$ and $CO_3^{2-}$. Under this assumption, oxygen isotope fractionation between $H_2O$ and DIC depends - in addition to temperature - on the pH of the solution because the relative contribution of individual carbonate species to precipitated calcite depends on the pH of the solution. This effect may also depend on different calcite precipitation reactions expressed by the PWP-equation (Plummer-Wigley-Parkhurst) (Plummer et al., 1978), which also depend on the pH of the solution.

In agreement with the conclusions of Coplen et al. (2007), Daëron et al. (2019) concluded that most natural cave carbonates, as well as most calcite samples obtained from laboratory precipitation experiments (e.g., Kim and O'Neil, 1997), are not precipitated in oxygen isotope equilibrium with water because of the comparably fast precipitation rates. They calculated an empirical temperature-dependent fractionation equation based on the analyses of extremely slow-growing calcites from Corchia Cave, Italy, and Devil's Hole, assuming that these samples have been precipitated in oxygen isotope equilibrium with water (DePaolo, 2011; Watkins et al., 2014). The calculated relationship for $^{18}\alpha_{calcite/H2O}$ express an apparent equilibrium limit. In contrast, the relationship of Kim and O'Neil (1997) likely expresses a kinetic limit for the oxygen isotope fractionation between calcite and water (Daëron et al., 2019) (Fig. 1). The equilibrium-to-kinetic range would be consistent with many observed oxygen isotope fractionation values obtained from speleothem calcite samples (Fig. 1). However, further studies have to test whether these temperature-relationships for oxygen isotope fractionation are truly the limits for cave carbonates. This suggests that the observed variability of temperature-dependent oxygen isotope fractionation is provoked by a combination of growth-rate related kinetic effects (Daëron et al., 2019; DePaolo, 2011; Dietzel et al., 2009; Hansen et al., 2019; Watkins et al., 2014) and oxygen isotope disequilibrium effects during the precipitation of secondary calcite on the speleothem surface or the watch glass, respectively (Deininger et al., 2012; Hansen et al., 2019; Mühlinghaus et al., 2009; Riechelmann et al., 2013). This may explain the variability of empirically determined oxygen isotope fractionation values (Fig. 1), especially for single cave systems, where the degree of kinetic and disequilibrium isotope effects may vary substantially (e.g. Affek et al., 2014).

All of the aforementioned studies assume that all DIC species are in oxygen isotope equilibrium with the drip water during secondary calcite precipitation on a speleothem surface or a watch glass, hence reporting ‘empirical’ oxygen isotope fractionation values. In the following, we argue that this assumption is violated in cave studies in case of prior calcite precipitation (PCP). PCP is calcite precipitation before the cave drip water drips on the surface of a stalagmite, including for instance calcite precipitation in karst fractures or on stalactite surfaces (e.g. Bernal et al., 2016; Cruz et al., 2007; Karmann et al., 2007; Millo et al., 2017; Riechelmann et al., 2011; Stoll et al., 2012).

## 2. The effect of PCP on primary and secondary calcite $\delta^{18}O$ values – the conceptual PCP-framework

The (equilibrium) isotope fractionation factor, $\alpha_{B/A}$, between two substances (in chemical and isotope equilibrium) (A <-> B) is defined as the ratio of the two isotope ratios, R, of the two substances $R_A$ and $R_B$: $\alpha_{B/A} = R_B/R_A$ (e.g., Mook, 2006). Thus, the equilibrium oxygen isotope fractionation factor between calcite and water, $^{18}\alpha_{calcite/H2O}$, is defined as $^{18}\alpha_{calcite/H2O} = {}^{18}R_{calcite}/{}^{18}R_{H2O}$. It describes the oxygen isotope fractionation from water to calcite (speleothem) resulting from multiple oxygen isotope fractionation processes in the $CO_2$-$H_2O$-$CaCO_3$ system during calcite precipitation and is positive in the range of relevant cave air temperatures (Fig. 1). Therefore, calcite has always a more positive $\delta^{18}O$ value ($\delta^{18}O_{calcite}$) compared to the $\delta^{18}O$ value of water. In the following, we only refer to $^{18}\alpha_{calcite/H2O}$ for equilibrium oxygen isotope fractionation but use the term ‘oxygen isotope fractionation’ for kinetic isotope fractionation. For speleothems, the oxygen isotope fractionation between calcite and water can be calculated by the ratio of the $^{18}R_{calcite}$ value of a recently precipitated secondary calcite sample (e.g., a speleothem or calcite on a watch glass) and the corresponding $^{18}R_{H2O}$ value of the drip water. It is important to consider that the calculated oxygen isotope fractionation factor will only correspond to the $\delta^{18}O_{H2O}$ value, if all chemical species of the $CO_2$-$H_2O$-$CaCO_3$ system (i.e., $CO^*_2$, $HCO_3^-$, $CO3^{2-}$) are in oxygen isotope equilibrium with the water. This assumption is generally used in theoretical modelling approaches (Deininger et al., 2012; Dreybrodt, 2008; Dreybrodt and Scholz, 2011; Mühlinghaus et al., 2007, 2009; Romanov et al., 2008; Scholz et al., 2009), experimental studies (Day and Henderson, 2011; Dietzel et al., 2009; Hansen et al., 2019; Polag et al., 2010; Wiedner et al., 2008) and in-situ cave studies (Daëron et al., 2019; Feng et al., 2012a; Feng et al., 2014; Johnston et al., 2013; Riechelmann et al., 2013; Tremaine et al., 2011). In the following, we argue that the assumption of oxygen isotope equilibrium between DIC and $H_2O$ is violated for cave drip water when PCP has occurred, which would consequently affect the calculated oxygen isotope fractionation values between calcite and water. To estimate the potential impact of this effect, we use the results of inorganic calcite precipitation experiments under cave-analogue conditions published by Hansen et al. (2019) and the proxy system model ISOLUTION (Deininger and Scholz, 2019).

## 2.1. Evolution of $\delta^{18}O$ and $\delta^{13}C$ values in inorganic calcite precipitation experiments

The following paragraph is a summary of the experimental setup for the inorganic calcite precipitation experiments and selected results of Hansen et al. (2019), and we refer to this study for further details. The inorganic calcite precipitation experiments were performed in a climate box equipped with gloves and an air-lock allowing to adjust for different experimental boundary conditions and take samples from the solutions and the gaseous $CO_2$ without contaminating the box's atmosphere. For the experiments, different $CO_2$-$H_2O$-$CaCO_3$ solutions were pumped onto inclined glass or marble plates and all 'environmental parameters', such as the $Ca^{2+}$ concentration of the solution, air temperature (T), relative humidity (rH), air $pCO_2$ as well as the isotope composition of the air $CO_2$ inside the box, were controlled, logged and adjusted to a natural cave environment. Caused by gravitation, the solution flowed down the plates as a thin solution film (ca. 0.1 mm in thickness), and $CaCO_3$ was progressively precipitated along the flow path (Fig. 2a). The experiments were performed at 10, 20 and 30 °C and $pCO_2$ of 1000 and 3000 ppmV. The initial $Ca^{2+}$ concentration of different solutions was 2, 3 and 5 mmol/L. It is important to note that the experiments were divided into two sub-sets of experiments. The first setup, performed on the marble plates, was used to study the temporal evolution of the carbonate chemistry as well as of the $\delta^{18}O$ and $\delta^{13}C$ values of the DIC of the solution along the flow path. The solution was collected at the end of the plate in an air-lock after traveling different distances representing different residence times on the marble plate. The isotope compositions of the DIC samples were preserved by quantitatively precipitating the DIC as $SrCO_3$. For the second part of the experiments, the marble plates were replaced by glass plates, and the remaining solution of a DIC experiment was used. The solution flowed down on an inclined borosilicate glass plate for about four days, and $CaCO_3$ was precipitated along the flow path at exactly the same boundary conditions as for the DIC experiments. Once an experiment was completed, the glass plate was removed from the climate box, and $CaCO_3$ samples were scratched off the glass plate at similar residence times (i.e., the same distances of flow) as in the DIC experiments for analyses of $CaCO_3$ $\delta^{18}O$ and $\delta^{13}C$. This sampling strategy allows combining the results obtained from the DIC and the $CaCO_3$ experiments and, for the first time, to directly study oxygen and carbon isotope fractionation processes between all involved species during calcite precipitation (speleothem formation) from a thin solution layer.

<<< Figure 2 >>>

For the development of the conceptual PCP-framework, we evaluate the results of the $CaCO_3$ precipitation experiments conducted using a 5 mmol/l $CaCO_2$-$CO_2$-$H_2O$ solution, at an air temperature of 20 °C and atmospheric $pCO_2$ values of 1000 and 3000 ppm, respectively (Fig. 2). The $Ca^{2+}$ concentration of the solution progressively decreases with increasing residence time on the plates (Fig. 2a), indicating continuous precipitation of $CaCO_3$ from the solution film. The precipitation rate is higher for the 1000 ppm experiment (Fig. 2a). As a consequence of progressive precipitation of $CaCO_3$ along the flow path and the associated isotope fractionation processes, the $\delta^{18}O$ and $\delta^{13}C$ values of both the precipitated $CaCO_3$ (Figs. 2b and d) and the DIC (Fig. 2c and e) increase. The $\delta^{18}O$ value of the water, in contrast, is constant during the experiments and not affected by the precipitation of $CaCO_3$ (Hansen et al., 2019).

The largest enrichment in $CaCO_3$ $\delta^{13}C$ values in these two experiments is approximately 7 ‰ and is observed for the longest distance of flow or the longest residence time on the plate (Fig. 2d). The largest enrichment in the $\delta^{13}C$ values of the DIC is ca. 7.5 ‰ and ca. 6 ‰ for the 1000 and 3000 ppm experiment, respectively, and again observed for the longest residence time on the plate (Fig. 2e). The enrichment in $\delta^{18}O$ values is generally lower compared to the enrichment in $\delta^{13}C$ values but also occurs at the longest residence time of the solution on the plates. It is about 1.9 ‰ for both $CaCO_3$ experiments (Fig. 2b) and approximately 2 ‰ and 1.4 ‰ for the 1000 and 3000 ppm DIC experiments, respectively (Fig. 2c). The temporal evolution of the $\delta^{18}O$ and $\delta^{13}C$ values can be explained using an isotope-enabled mass balance model, and we refer to Hansen et al. (2019) for a detailed discussion on this aspect. These experiments demonstrate that the DIC $\delta^{18}O$ and $\delta^{13}C$ values are increasing as a result of progressive precipitation of calcite. Furthermore, the evolution of $\delta^{18}O$ and $\delta^{13}C$ values of the precipitated $CaCO_3$ is strongly coupled with the evolution of $\delta^{18}O$ and $\delta^{13}C$ values of the DIC. Because the $\delta^{18}O_{H2O}$ of the solution is constant, the progressive precipitation of calcite results in the observed oxygen isotope disequilibrium of the DIC and $CaCO_3$ $\delta^{18}O$ values with respect to water.

The calcite precipitation experiments can be considered an experimental analogue for PCP, and the plates are analogues for, e.g., a stalactite hanging above a stalagmite. Thus, the results from the calcite precipitation experiments of Hansen et al. (2019)

facilitate to better constrain the effect of PCP on $\delta^{18}O$ values of speleothem calcite and the calculation of the oxygen isotope fractionation between calcite and water.

**2.2 Modelling of the evolution of $\delta^{18}O$ values in $HCO_3^-$**

To complement the experimental results, particularly on longer time scales (hours to days), we use the ISOLUTION proxy system model (PSM) (Deininger et al., 2012; Deininger and Scholz, 2019). ISOLUTION is an isotope-enabled mass balance model designed for thin solution films and describes the evolution of the $\delta^{18}O$ (and $\delta^{13}C$, not shown here) values of $HCO_3^-$ in the solution as well as that of the precipitated calcite during calcite precipitation (Deininger et al., 2012; Scholz et al., 2009). The focus is on $HCO_3^-$, because for solutions with pH values between 8.1 and 8.7, which is the case for most natural cave waters, the $\delta^{18}O$ value of the precipitated calcite mainly originates from the drip water $HCO_3^-$ (>97.5 % of the DIC). ISOLUTION accounts for (i) oxygen isotope fractionation effects caused by the conversion of $HCO_3^-$ and $Ca^{2+}$ into $CO_2$, $H_2O$ and calcite ($CaCO_3$) (i.e., calcite precipitation), (ii) the oxygen isotope exchange between $H_2O$ and $HCO_3^-$ and (iii) the evaporation of $H_2O$ from the solution and the condensation of vapour of the cave air (Deininger et al., 2012; Deininger and Scholz, 2019). In the following discussion, two processes mainly modify the evolution of the $\delta^{18}O$ value of $HCO_3^-$: the oxygen isotope fractionation processes during calcite precipitation ($Ca^{2+}+2HCO_3^- \rightarrow CaCO_3+CO_2+H_2O$) and the oxygen isotope exchange between $H_2O$ and $HCO_3^-$. Evaporation of $H_2O$ can be ignored, because the relative humidity is above 90% in most caves, and evaporation effects are negligible in such cases (Deininger et al., 2012; Dreybrodt and Deininger, 2014). Therefore, we assume for the following discussion that the relative humidity is 100 % (i.e. no evaporation) and that the vapour of the cave air is in oxygen isotope equilibrium with the $H_2O$ of the solution. Furthermore, the thickness (d) of the solution film on the speleothem surface is 0.1 mm – a good approximation for solution films on speleothem surfaces (Dreybrodt, 1980) – and we used the related calcite precipitation rate constant, $\lambda_P$, estimated by Baker et al. (1998) for the following calculations. The calcite precipitation time constant, $\tau_P$, can then be calculated for a given cave air temperature by the ratio between d and $\lambda_P$ (Kaufmann, 2003) (see also Deininger and Scholz, 2019 for details about ISOLUTION).

For the following ISOLUTION calculation (Fig. 3), we assume that carbonate dissolution in the karst occurs under open-system conditions at a $pCO_{2.karst}$ of 5000 ppm and that all carbonate species are in chemical and isotope equilibrium before PCP starts to occur. The oxygen isotope ratio of the infiltrated water, percolating through the karst, has a $\delta^{18}O_{H2O}$ value of -10 ‰ with respect to V-SMOW; the $pCO_2$ value of the cave air ($pCO_{2.cave}$) during PCP is 2000 ppm, and the ambient air temperature is 10 °C. For another set of ISOLUTION calculations (Fig. 4), we varied the $pCO_{2.karst}$ value using 3000 ppm and 10,000 ppm, to investigate the effect of super-saturation with respect to calcite on the oxygen isotope disequilibrium between $H_2O$ and $HCO_3^-$ and between $H_2O$ and calcite. The latter can be inferred for this theoretical case from the oxygen isotope disequilibrium between $H_2O$ and $HCO_3^-$.

<<< Figure 3 >>>

Consistently with the experimental results (Fig. 2), the $\delta^{18}O$ values of $HCO_3^-$ (and the calcite) increase when calcite precipitation begins. As a result, $\delta^{18}O$ values of $HCO_3^-$ deviate from the equilibrium $\delta^{18}O$ value with respect to $H_2O$ (Fig. 3). The oxygen isotope disequilibrium of the $HCO_3^-$ (and calcite) with respect to $H_2O$ increases with an increasing difference between $pCO_{2.karst}$ and $pCO_{2.cave}$, i.e., with increasing super-saturation with respect to calcite (Fig. 4). After a maximum $\delta^{18}O$ value of $HCO_3^-$ has been reached (within several hundreds of seconds), the $\delta^{18}O$ value begins to decrease until $HCO_3^-$ is again in oxygen isotope equilibrium with $H_2O$. At an air temperature of 10 °C, the oxygen isotope re-equilibration takes about 141,200 s (c. 1.6 days) (Figs. 3 and 4).

<<< Figure 4 >>>

The chemical reactions of calcite precipitation (conversion of $Ca^{2+}$ and $HCO_3^-$ into $CaCO_3$, $H_2O$ and $CO_2$) are accompanied by oxygen (and carbon) isotope fractionation effects, which, in combination, cause the observed increase in $\delta^{18}O$ (and $\delta^{13}C$, not shown) values of $HCO_3^-$ (Scholz et al., 2009) (Figs. 3 and 4). As a result of the higher $Ca^{2+}$ (and $HCO_3^-$) concentration in the solution for higher $pCO_{2.karst}$ values and a constant $pCO_{2.cave}$ value, the precipitation rate increases, thereby increasing the oxygen isotope disequilibrium between $HCO_3^-$ and $H_2O$ (Fig. 4) (Deininger et al., 2012; Deininger and Scholz, 2019). Simultaneously, the continuous oxygen isotope exchange between $HCO_3^-$ and $H_2O$ balances the oxygen isotope disequilibrium between $HCO_3^-$ and $H_2O$ caused by calcite precipitation (Scholz et al., 2009). Therefore, the evolution of the $\delta^{18}O$ value of the dissolved $HCO_3^-$ during calcite precipitation and that of the precipitated calcite depends on the competitive oxygen isotope effects caused by calcite precipitation and oxygen isotope exchange between $HCO_3^-$ and $H_2O$. The temporal evolution of the $\delta^{18}O$ value of the $HCO_3^-$ is governed by the chemical disequilibrium of the solution with respect to the $pCO_{2.cave}$ and the time constants for the calcite precipitation, $\tau_P$, and for the oxygen isotope exchange, $\tau_B$. These are, in turn, related to temperature. According to Dreybrodt and Scholz (2011), $\tau_B$ is on the order of 125,000 s at 0 °C; 35,300 s at 10 °C and 10,600 s at 20 °C (Table 1). For d = 0.1 mm, $\tau_P$ varies from 338 to 2,064 s in the same temperature range (Table 1) and is in agreement with the experimentally observed values for $\tau_P$ (see Table 1 in Hansen et al. 2019).

<<< Table 1 >>>

Because $\tau_B$ is at least two orders of magnitude larger than $\tau_P$ (i.e., the oxygen isotope exchange between $HCO_3^-$ and water is slower compared to calcite precipitation), oxygen isotope effects related to calcite precipitation predominate the effect of oxygen isotope exchange at the beginning of the calcite precipitation and result in the observed (Fig. 2) and modelled (Figs. 3 and 4) increase in the $\delta^{18}O$ and $\delta^{13}C$ values of $HCO_3^-$ and $CaCO_3$ ($\delta^{13}C$ values are only shown for the experiments). When the solution is close to or in chemical equilibrium with the surrounding atmosphere again, the net calcite precipitation rate is reduced, and the accompanying effects of $HCO_3^-$-$CaCO_3$ isotope fractionation become less dominant. Oxygen isotope exchange between $HCO_3^-$ and $H_2O$, in contrast, is a continuous process and results in nearly complete re-equilibration (98 %) between the oxygen isotope composition of $HCO_3^-$ and $H_2O$ after about four times of $\tau_B$ (Figs. 3 and 4) (Scholz et al., 2009). Thus, the time for re-establishing the oxygen isotope equilibrium between $HCO_3^-$ (DIC) and $H_2O$ is 5.6 days, 1.6 days and 0.5 days at temperatures of 0 °C, 10 °C and 20 °C, respectively (Table 1). Therefore, in case of PCP before the drip water reaches the location of an actively growing stalagmite, the drip water would only be in oxygen isotope equilibrium if enough time passed between PCP and secondary calcite precipitation on a stalagmite surface or on a watch glass. This implies that the residence/travel time of cave water has important consequences for the $\delta^{18}O$ values of speleothem calcite and the determination of $^{18}\alpha_{calcite/H2O}$ via cave carbonates and drip water.

## 3. Discussion

### 3.1 The effect of PCP on secondary calcite $\delta^{18}O$ values

Based on the results of sections 2.1 and 2.2, we investigate the potential effect of PCP on the apparent oxygen isotope fractionation between secondary calcite in caves and cave drip water, calculated from the $\delta^{18}O$ values of recent secondary calcite and the corresponding drip water samples: $\alpha_{calcite/water}=(1000+\delta^{18}O_{calcite})/(1000+\delta^{18}O_{dripwater})$. As explained in the previous sections, PCP disturbs the oxygen isotope equilibrium between $HCO_3^-$ (DIC) and $H_2O$, resulting in more positive $\delta^{18}O_{calcite}$ values of secondary calcite while changes in $\delta^{18}O_{water}$ are insignificant. Therefore, the oxygen isotope fractionation factor appears larger than it actually is if calculated from such a set of secondary calcite and drip water samples. In natural settings, it is essential to assess where along the flow path of the percolating water PCP takes place, because the time available for oxygen isotope re-equilibration between $HCO_3^-$ and drip water can vary substantially if PCP occurs in the karst or the cave.

If PCP occurred in the karst (Fig. 5a), the solution would further percolate through the host rock after PCP terminated (Fig. 5a), and until the solution enters the cave, the PCP generated oxygen isotope disequilibrium between $HCO_3^-$ and $H_2O$ may be balanced by the oxygen isotope exchange between $HCO_3^-$ (DIC) and $H_2O$ (Section 2.2). If the time interval between the termination of PCP in the karst and the moment when the solution drips onto the top of a stalagmite is too short for complete oxygen isotope re-equilibration between $HCO_3^-$ (DIC) and $H_2O$, the $\delta^{18}O$ value of the $HCO_3^-$ would be still in oxygen isotope disequilibrium with respect to water (Fig. 4). For example, when a solution equilibrated to a $pCO_{2.karst}$ of 10,000 ppm enters a cavity with a $pCO_{2.cave}$ of 2000 ppm at a temperature of 10 °C, the $\delta^{18}O$ value of the $HCO_3^-$ would be more positive (by about 0.6 ‰) if this solution entered the cave 12 hours after the termination of PCP compared to oxygen isotope equilibrium conditions between water and $HCO_3^-$ (Fig. 4). As a consequence, the $\delta^{18}O$ value of the secondary calcite precipitated on the stalagmite or watch glass would also be enriched by 0.6 ‰ compared to the expected $\delta^{18}O_{calcite}$ value when oxygen isotope equilibrium is established between $HCO_3^-$ (DIC) and water. In natural cave environments, it is often difficult to assess whether PCP has occurred in the karst because the isotopic and geochemical evolution of the solution in the karst cannot be monitored. It is even more difficult to determine the degree of PCP. The occurrence and degree of PCP in the karst can only be inferred indirectly using additional speleothem proxies, such as speleothem Mg/Ca and Sr/Ca ratios (Fairchild and Treble, 2009; Stoll et al., 2012). However, even then, it is not possible to determine how much time has passed since the termination of PCP.

In contrast, it is much easier to examine if PCP has occurred on the cave ceiling or stalactite surfaces because these are accessible for observation. For example, a wet stalactite surface and the formation of drops at the tip of a stalactite can be an indication. However, the visual examination of stalactites or the flow path along the cave ceiling does not allow to infer detailed information on the degree of PCP. If PCP occurred on the cave ceiling or an actively growing stalactite above a stalagmite or watch glass (Fig. 5b), the PCP-effect on secondary calcite $\delta^{18}O$ values may be more pronounced than when PCP takes place in the karst. This is because the residence time of the cave water on the cave ceiling or the stalactite after the termination of PCP is much shorter (or non-existent) than the time required for the oxygen isotope re-equilibration between $HCO_3^-$ (DIC) and water (e.g., 1.6 days at 10 °C, Table 1). The degree of PCP inside caves mainly depends on two parameters. Firstly, the residence time of the cave water on the cave ceiling or the stalactite surface, i.e., the time interval when calcite precipitates on the cave ceiling or the surface of a stalactite. Another crucial factor that controls the degree of PCP are the temperature-dependent chemical reaction rates of calcite precipitation. In general, for longer residence times on the cave ceiling or the surface on a stalactite and higher precipitation rates, the degree of PCP would be stronger. Therefore, the effect of PCP on speleothem calcite $\delta^{18}O$ values becomes stronger for longer drip intervals and larger stalactites, respectively. Furthermore, the precipitation rate and the degree of PCP increase with increasing cave air temperatures.

Beside these qualitative relationships, we theoretically estimate the degree of PCP for different boundary conditions (Table 2). These calculations refer to an upper limit for the degree of PCP along the stalactite surface that can be approximated using the information on the geometry of the stalactite (radius and length), the drip interval and the Stokes equation for gravity-driven flow (Short et al., 2005). The drip interval allows calculating the volumetric drip water flux on the entire stalactite surface, assuming that one droplet has a volume of 0.1 ml (≙ 0.1 $cm^3$) (e.g., Mühlinghaus et al., 2007). Then, the mean flow rate and the thickness of the fluid film can be calculated (Eqs. (4) and (5) in Short et al., 2005). This, in turn, allows us to estimate the residence time of the cave water on the stalactite surface and, depending on the thickness of the fluid film, the calcite precipitation rate at a given cave air temperature (see Table 2 for details about these calculations). We emphasise here that the residence time only refers to the time interval that the cave water flows down on the stalactite surface as a thin film and does not include the time required to form a drop, which then hangs at the tip of the stalactite. PCP at the tip of the stalactite during the formation of a drop would only further increase the degree of PCP. The qualitative estimate is confirmed by the calculations (Table 2), which indicate that the degree of PCP increases with longer residence times caused either by longer drip intervals or larger stalactites. For a 37 cm long stalactite with a base radius of 2.5 cm at a cave air temperature of 10 °C, the degree of PCP can be as high as 96.7 % when the drip interval is very long (1800 s). For shorter drip intervals (120 s), it would still be considerably high (42.8 %).

It should be noted that the assumption that the cave water flows down evenly as a thin film on the entire stalactite surface may not apply to natural cave environments. This is because cave water tends to flow down the surface of a stalactite as small ‘rivulets’, particularly for stalactites of larger dimension. When the effect of rivulets is accounted for in the calculations, the thickness and the flow rate of the fluid film scales with a factor of $(1/y)^{1/3}$ and $(1/y)^{2/3}$, where the factor 1/y is the fraction of the stalactite surface where the cave water flows down. The scaling parameters are derived from Eq. (3)

(the volumetric drip flux) in Short et al. (2005), by scaling the circumference of the symmetric stalactite. Therefore, the thickness of the fluid film would increase and the residence time of the fluid film on the stalactite surface would decrease, consequently decreasing the degree of PCP. The result of this considerations is that if the cave water flows down on 1/10$^{th}$ of the stalactite surface, and the degree of PCP decreases to 51.9 % (1800 s) and 11.3 % (120 s) compared to 96.7 % and 42.8 % in the example above. The degree of PCP decreases even further to 36.9 % (1800 s) and 7.3 % (120 s) if the cave water flows down on 1/20$^{th}$ of the stalactite surface. Therefore, the approximation of the degree of PCP presented in Table 2 is an upper limit describing an even film on the entire stalactite surface. These estimates indicate that the degree of PCP is probably small in case of small stalactites and short drip intervals, in particular when the cave water flows down the surface of a stalactite as small rivulets. However, the degree of PCP can be considerably large for large stalactites and long drip intervals. Note that if the chemical equilibrium has already been reached during PCP inside the cave (e.g., on a stalactite surface), the net precipitation rate of secondary calcite would be very small. This is likely to be different when PCP occurs in the karst, where the $pCO_2$ value of air captured in fractures might be still higher than the $pCO_2$ of the cave air. Thus, secondary calcite precipitation in the cave would be still possible. In combination with proxy system models, such as ISOLUTION (Deininger and Scholz, 2019), our calculations can be used to estimate the PCP-induced $\delta^{18}O$ bias in secondary calcite, also accounting for effects caused by the supersaturation of the drip water and temperature-dependent oxygen isotope fractionation.

In summary, the degree of PCP and the time interval between the termination of PCP and the precipitation of secondary calcite determine whether the $\delta^{18}O$ value of the secondary calcite is precipitated in oxygen isotope equilibrium with water or not. If the secondary calcite is not precipitated in oxygen isotope equilibrium with water, the $\delta^{18}O_{calcite}$ value would be more positive, because of the corresponding oxygen isotope disequilibrium between $HCO_3^-$ (DIC) and $H_2O$. This is likely to be the case when PCP occurs in the cave, e.g., on a stalactite surface, rather than when PCP occurs in the karst. If the oxygen isotope fractionation was calculated from the $\delta^{18}O$ values of this secondary calcite and the corresponding $H_2O$, the speleothem oxygen isotope fractionation factor between calcite and water would be overestimated. The overestimation would increase with an increasing degree of PCP and a shorter time interval between the termination of PCP and the moment when secondary calcite precipitation begins (Fig. 5b). Therefore, when attempting to calculate an empirical oxygen isotope fractionation factor from cave water and secondary calcite samples in caves, either from speleothem surfaces or collected on watch glasses, it is crucial to know or estimate if and where PCP occurred. Such an indicator is, for example, when a stalactite grows directly above a sampled stalagmite or watch glass. In such cases, the geometry of the stalactite, the dimension of the cave water flowing down on the stalactite surface and the drip interval are important parameters that control the degree of PCP on the stalactite.

**3.2 Analytical limitations for the calculation of the oxygen isotope fractionation**

The oxygen isotope equilibrium between DIC ($HCO_3^-$) and $H_2O$ is disturbed when PCP occurs, which results in an apparently larger cave water-calcite oxygen isotope fractionation factor than under conditions of oxygen isotopic equilibrium (Figure 5). This PCP-related effect is caused by the increase of the $\delta^{18}O$ values of $HCO_3^-$ during precipitation of $CaCO_3$, while the $\delta^{18}O$ of the water remains unchanged during PCP. This is because the number of $H_2O$ molecules is much larger than that of all DIC ($HCO_3^-$) molecules (Scholz et al., 2009) and has been confirmed in the $CaCO_3$-precipitation experiments of Hansen et al. (2019). Therefore, if the cave drip water and the respective (modern) secondary calcite deposited on a speleothem or watch glass are sampled from a drip site that has experienced PCP and oxygen isotope equilibrium with the cave drip water has not been re-established, the calcite $\delta^{18}O$ value will be more positive compared to the expected equilibrium calcite $\delta^{18}O$ value with respect to water (Figure 4).

For calculating the oxygen isotope fractionation factor between calcite and water, a recently precipitated calcite sample on a speleothem surface or watch glass and the corresponding drip water sample are required. In the case of PCP, the period between the termination of PCP and the moment of drip water sampling is important (Section 3.1). In the following, we assume an ideal sampling of the drip water, in which no $CO_2$ exchange with the atmosphere, no evaporation of $H_2O$ occurs, and no air is captured inside the sampling vessel. Hence, the only process potentially affecting the $\delta^{18}O$ value of the dissolved $HCO_3^-$ is oxygen isotope exchange between $HCO_3^-$ and $H_2O$. If no PCP occurred, the $\delta^{18}O$ value of the $HCO_3^-$ is in oxygen isotope equilibrium with the water and, if kinetic and disequilibrium isotope fractionation during precipitation of $CaCO_3$ are not present (ignored), the calcite $\delta^{18}O$ value would reflect the equilibrium $\delta^{18}O$ value with respect to the $\delta^{18}O$ value of $H_2O$ (Fig. 4). Therefore, if the oxygen isotope fractionation factor was calculated from the $H_2O$ and calcite $\delta^{18}O$ values, it would correspond to the true equilibrium oxygen isotope fractionation factor between $H_2O$ and calcite ($^{18}\alpha_{calcite/H2O}$) at the given cave air temperature. We note again that in natural cave systems, the assumption of no kinetic and disequilibrium isotope processes during the speleothem formation is likely not met (Daëron et al., 2019), also shown by the laboratory calcite precipitation experiments of Hansen et al. (2019).

If PCP occurred and the solution was sampled before oxygen isotope re-equilibration between $HCO_3^-$ and $H_2O$, the $\delta^{18}O$ value of $HCO_3^-$ would be enriched in $^{18}O$ (more positive) compared to the case when no PCP occurred. This implies that the secondary calcite that is precipitated from this solution is also enriched in $^{18}O$ and the corresponding $\delta^{18}O$ value would be more positive compared to the calcite $\delta^{18}O$ values when $HCO_3^-$ is in oxygen isotope equilibrium with $H_2O$ (Fig. 4). We note that the $\delta^{18}O$ value of $H_2O$ is constant and not altered by PCP.

In principle, there are two methods to analyse the water and calcite samples after sampling, the conventional approach where the drip water and the secondary calcite sample are analysed in the laboratory and the 'in-situ' approach where the DIC is precipitated shortly after the sampling in the cave. For the conventional approach, the calculated value of the oxygen isotope fractionation factor would be biased (i.e., too large). As outlined above, the time for the oxygen isotope re-equilibration between $HCO_3^-$ (DIC) and $H_2O$ is about 1.4 and 3.1 days at temperatures of 10 °C and 5 °C, respectively (Table 1). The time interval from the moment of sampling of the cave drip water until the measurement of the corresponding $\delta^{18}O$ value in the laboratory is likely much longer compared to the time required for the oxygen isotope re-equilibration because logistic reasons (transport of the sample, possible storage times in the lab) often lead to long waiting times (up to several months) before measurement. In addition, the most widely applied technique to measure the $\delta^{18}O$ value of water samples (e.g. Cruz et al., 2005; Pape et al., 2010; Spötl et al., 2005; Yonge et al., 1985) is the $CO_2$-equilibration method (Epstein and Mayeda, 1953), which is based on the oxygen isotope exchange reaction between DIC and $H_2O$. Therefore, even if the cave drip water sample had not reached oxygen isotope equilibrium at the time of stable isotope analysis, the disequilibrium offset would be erased by the $CO_2$-equilibration method. In practice, only the $\delta^{18}O$ value of $H_2O$ is inferred from the stable isotope measurements and not the $\delta^{18}O$ value of $HCO_3^-$ or DIC at the moment of sampling. The same is true if only the $\delta^{18}O$ value of the water would be measured by spectrometric techniques (Dublyansky et al., 2019). Hence, for both techniques, the $\delta^{18}O$ value of the analysed drip water does not reflect the $\delta^{18}O$ value of the $HCO_3^-$ (DIC) at the moment of drip water sampling. Thus, if the oxygen isotope fractionation factor was calculated from such a set of secondary calcite and drip water $\delta^{18}O$ values, the inferred oxygen isotope fractionation factor is more positive and would overestimate the oxygen isotope fractionation between calcite and water. In addition, changing temperatures during the transport out of the cave to the laboratory and subsequent storage affect the $\delta^{18}O$ value of the DIC probably several times at different magnitudes, because the relative portions of the DIC-species and the oxygen isotope fractionation between the DIC-species change with temperature.

Therefore, the only technique to directly investigate the $\delta^{18}O$ value of $HCO_3^-$ (DIC) at the moment of drip water sampling and to quantify the potential effect of PCP on $HCO_3^-$ (DIC) would be to quantitatively precipitate the DIC as $SrCO_3$ or $BaCO_3$ inside the cave (in-situ approach), such as conducted in laboratory experiments (Beck et al., 2005; Dreybrodt et al., 2016; Hansen et al., 2017; Hansen et al., 2019). Such samples can easily be stored without further oxygen isotope alteration and later be analysed like conventional carbonate samples in the laboratory (e.g., Hansen et al. 2019).

To conclude, the outlined arguments demonstrate that PCP results in $^{18}O$ enriched secondary calcite (increased calcite $\delta^{18}O$ value) if the time interval between the termination of PCP and the precipitation of secondary calcite is too short compared to the time required for the oxygen isotope re-equilibration between $HCO_3^-$ and $H_2O$. This effect will result in an overestimated oxygen isotope fractionation factor between calcite and water because the PCP effect on the $\delta^{18}O$ value of $HCO_3^-$ (DIC) cannot be revealed with currently applied techniques. The result may be erroneously interpreted as increased kinetic and isotope disequilibrium effects during calcite precipitation.

### 3.3 Variability of cave-based oxygen isotope fractionation factors

It is of great interest whether the observed variability of cave-based isotope fractionation factors (Fig. 1) can be explained by the discussed effects of PCP on the inferred oxygen isotope fractionation factor. We emphasise again that we ignore kinetic and disequilibrium isotope effects at this point of the discussion. Inorganic calcite precipitation experiments are appropriate analogues to estimate the effect of PCP on the calculation of the oxygen isotope fraction factor and demonstrated that the increase in calcite $\delta^{18}O$ values could be up to 1.5 ‰ (Polag et al., 2010; Wiedner et al., 2008) and 3.5 ‰ (Fig. 2) (Hansen et al., 2019), respectively, depending on temperature, super-saturation of the solution with respect to calcite and the duration of calcite precipitation. If PCP is not considered in the calculation of the oxygen isotope fractionation factor between secondary calcite and drip water samples from caves, the bias in the oxygen isotope fractionation factor can be in the same range as observed for the laboratory calcite experiments. From the compilation of studies illustrated in Fig. 1 only Johnson et al. (2006) state that PCP may have affected the studied drip sites and considers PCP to explain the observed variations in the cave drip water $Ca^{2+}$ concentration. Genty (2008) chose different places beneath stalactites for seepage water analysis. The calculated oxygen isotope fractionation factors between calcite and water of Tremaine et al. (2011) are likely also influenced by PCP to some degree because they demonstrated that PCP affects trace element concentrations at their study site (Tremaine and Froelich, 2013). Thus, for these studies, it is very likely that PCP affected the $\delta^{18}O$ value of calcite and consequently resulted in biased calcite/water oxygen isotope fractionation factors. Other studies mention disequilibrium isotope effects (Fuller et al., 2008; Mickler et al., 2004; Mickler et al., 2006; Plagnes et al., 2002) that can also cause a biased calcite/water oxygen isotope fractionation factor. However, except for (Johnson et al., 2006) and (Tremaine and Froelich, 2013), none of the other studies tested if PCP affected their drip sites. Hence, the values for oxygen isotope fractionation between calcite and water derived from these studies may not reflect 'equilibrium' $^{18}\alpha_{calcite/H2O}$ values and can overestimate the oxygen isotope fractionation between calcite and $H_2O$. Therefore, PCP cannot be excluded to be the reason for the observed variability in $^{18}\alpha_{calcite/H2O}$ values calculated from speleothem calcite and drip water samples (Fig. 1), also other effects such as kinetic isotope effects can provoke similar biases (section 3.4).

### 3.4 The effect of kinetic isotope effects on the oxygen isotope fractionation factor derived by speleothems

So far, we considered isotope fractionation processes occurring during PCP. The very same processes occur during precipitation of secondary calcite on the surface of stalagmites, flowstones or watch glasses. As during PCP, progressive precipitation of secondary calcite may disturb the oxygen isotope equilibrium between DIC ($HCO_3^-$) and water resulting in increased secondary calcite $\delta^{18}O$ values (compare section 2). The effect will depend on the degree of super-saturation of the drip water and the drip interval (Deininger et al., 2012; Deininger and Scholz, 2019; Mühlinghaus et al., 2009). Therefore, precipitation of secondary calcite on the surface of a speleothem will either cause isotopic disequilibrium between the DIC and the drip water or further amplify the effect of PCP, especially for long drip intervals. In both cases, the oxygen isotope fractionation factor calculated from secondary calcite and the corresponding drip water overestimates the oxygen isotope fraction between calcite and water.

Another aspect that has been ignored in the discussion so far is kinetic isotope fractionation during calcite precipitation. The assumption of isotope equilibrium is not valid for both inorganic precipitation experiments and speleothem growth (Daëron et al., 2019; Hansen et al., 2019). Kinetic isotope effects include, for example, the rate-dependence of the oxygen (and carbon) isotope fractionation factor between calcite and $HCO_3^-$. Therefore, in addition to the potential effects of PCP and secondary calcite precipitation, these kinetic isotope effects need to be considered when calculating 'equilibrium' or 'kinetic' oxygen isotope fractionation factors from secondary calcite $\delta^{18}O$ values and corresponding cave drip water. Daëron et al. (2019) outlined that the kinetic effects crucially depend on the net precipitation rate during secondary calcite formation (DePaolo, 2011; Hansen et al., 2019; Watkins et al., 2014). Therefore, depending on the net precipitation rate during calcite precipitation, 'equilibrium', 'kinetic' as well as intermediate (i.e., between equilibrium and kinetic) calcite/water oxygen isotope fractionation factors could be obtained, and kinetic calcite/water oxygen isotope fractionation factors would always be more negative than the corresponding equilibrium $^{18}\alpha_{calcite/H2O}$. This interplay of kinetic isotope effects and isotope disequilibrium effects resulting from progressive calcite precipitation during PCP and precipitation of speleothem calcite further complicates the determination of calcite/water oxygen isotope fractionation factors based on speleothem calcite. This may also explain the large variability observed in calculated calcite/water oxygen isotope fraction factors from cave studies, particularly for those obtained from a single cave system (Fig. 1).

## 4. Conclusions

In summary, the answer to our question, '*Are oxygen isotope fractionation factors between calcite and water derived from speleothems systematically biased due to prior calcite precipitation (PCP)?'*, is – for many caves and speleothem samples – probably yes, but this conclusion needs to be verified by new techniques for each sampling site. In detail:

1) The conceptual idea that PCP affects drip water DIC ($HCO_3^-$) $\delta^{18}O$ values resulting in oxygen isotope disequilibrium with respect to the $\delta^{18}O$ value of the drip water is an important process to explain the observed variability of oxygen isotope fractionation factors calculated from cave drip water and the corresponding secondary calcite $\delta^{18}O$ values as well as the offset to the $^{18}\alpha_{calcite/H2O}$ values derived from laboratory experiments, as, for instance, by Kim and O'Neil (1997) (Fig. 1).
2) The kinetic isotope fractionation processes and the disequilibrium effects between DIC ($HCO_3^-$) and $H_2O$ resulting from progressive precipitation of calcite are counter-acting effects. While increasing kinetic isotope effects tend to decrease the oxygen isotope fractionation between calcite and water (Daëron et al., 2019; DePaolo, 2011; Watkins et al., 2014), progressive precipitation of calcite increase the apparent oxygen isotope fractionation between calcite and water (Deininger et al., 2012; Deininger and Scholz, 2019; Mühlinghaus et al., 2009).

3) These processes provide an explanation for the observed variability of empirical calcite/water oxygen isotope fractionation derived from speleothem samples grown at the same cave temperature (Fig. 1), which may be caused by a combination of different degrees of kinetic and disequilibrium isotope effects.

4) Nevertheless, speleothem calcite has the potential to be used for the estimation of 'kinetic' or 'equilibrium' calcite/water oxygen isotope fractionation factors, if PCP and isotope disequilibrium processes could be accounted for in calculations of proxy system models.

5) Equilibrium $^{18}\alpha_{calcite/H2O}$ values may be obtained from drip sites with no stalactites above (no PCP), low super-saturation with respect to calcite (slow growth rates, i.e., small or no kinetic effects) and short drip intervals (low or negligible isotope disequilibrium effects during calcite precipitation). Drip sites that do not fulfil these criteria may only reflect local calcite-water oxygen isotope fraction substantially influenced by kinetic and isotope disequilibrium effects between DIC ($HCO_3^-$) and $H_2O$.

6) Considering that the residence time in the karst can be multiple times longer than the re-equilibration time of the drip water, it is likely that the cave drip water is in oxygen isotope equilibrium when it enters the cave after PCP has terminated in the karst. This is not the case when a stalactite grows above a stalagmite or watch glass experiment, where the time for re-equilibration with the water would be much longer compared to the residence time of the solution on the stalactite surface.

7) The effect of oxygen isotope disequilibrium between the DIC and the $H_2O$ may be estimated within cave monitoring programs and quantified using PSMs, such as ISOLUTION or KARSTOLUTION (Deininger et al., 2012; Deininger and Scholz, 2019; Treble et al., 2019).

Our conceptual PCP-framework but also other theoretical as well as experimental studies that investigated kinetic and disequilibrium isotope effects in speleothems demonstrate that multiple kinetic isotope processes occur at the same time which all modify $\delta^{18}O$ (and $\delta^{13}C$) values in speleothems (section 1). While these processes may prevent the use of speleothem $\delta^{18}O$ as a stable isotope thermometer, oxygen and carbon isotopes in speleothems allow versatile insights into climate and environmental changes. This is because the degree of kinetic and disequilibrium isotope effects in speleothems depend on environmental parameters such as temperature, drip interval, super-saturation of the drip water and cave ventilation. In addition, the variability of drip water $\delta^{18}O$ (and $\delta^{13}C$) values imprint changes of precipitation $\delta^{18}O$ (soil $CO_2$ $\delta^{13}C$), thus hydroclimate (vegetation) changes from above the cave. This renders the possibility for qualitative climate and environmental reconstructions using speleothem $\delta^{18}O$ (and $\delta^{13}C$). The development of inverse multi-proxy system models in the future may render the possibility for quantitative climate and environmental reconstructions.

**Acknowledgements**

MD thanks three anonymous reviewers, whose comments help to improve the manuscript. MD, MH, JF and DS acknowledge funding by the German Research Foundation (DFG) through grants DE 2398/3-1, HA8694/1-1, FO 809/6-1 and SCHO 1274/11-1.

**Research Data**

All experimental data shown in Fig. 2 and 5 are available as an Electronic Annex in Hansen et al. (2019) (doi.org/10.1016/j.chemgeo.2018.12.012). The MATLAB code for the ISOLUTION Proxy System Model is published by Deininger and Scholz (2019) and publicly available at doi.org/10.5038/1827-806X.48.1.2219

## Figure Caption

**Figure 1:** Compilation of $1000\ln(^{18}\alpha_{calcite/H2O})$ values that are derived from various speleothem studies. The shading areas indicate the range of derived $1000\ln(^{18}\alpha_{calcite/H2O})$ from the respective studies. The black solid and dashed lines depict the temperature relationship for $1000\ln(^{18}\alpha_{calcite/H2O})$ of Kim and O'Neil (1997) and Daëron et al. (2019), representing a kinetic and an equilibrium 'limit' for $1000\ln(^{18}\alpha_{calcite/H2O})$, respectively.

**Figure 2**: Compilation of the experimental results from Hansen et al. (2019) performed using a 5 mmol/l $CaCO_3$-$CO_2$-$H_2O$ solution at 20 °C and 1000 and 3000 ppm $CO_2$ with respect to the residence time of the solution on the plates. a) remaining fraction of $Ca^{2+}$ in the solution, the black triangles represent the experiment performed at 1000 ppm, the grey triangles represent the one performed at 3000 ppm $CO_2$ atmosphere; b) and d) shows the results for the stable isotope measurements of the $CaCO_3$ along the flow path as deviation Δ from the initial isotope ratio value [ $\Delta = \delta_{t(s)} - \delta_{initial}$ ] (i.e. 0 cm distance of flow, corresponding to the apex of a stalagmite); panel c) and e) show the results for the stable isotope measurements of the DIC as Δ. In panels b to e the red symbols represent the $\delta^{13}C$ and the blue symbols the $\delta^{18}O$ values.

**Figure 3:** Temporal evolution of the $\delta^{18}O$ value of $HCO_3^-$ (black line) and $H_2O$ (grey line) in the case of calcite precipitation as modelled using ISOLUTION. The parameters are: $pCO_{2.karst}$ = 5000 ppm, $pCO_{2.cave}$ = 2000 ppm, T 10 °C. The dashed line highlights the equilibrium $\delta^{18}O$ value of $HCO_3^-$ with respect to the $\delta^{18}O$ value of $H_2O$ (Kim and O'Neil, 1997). The difference between the dashed and the solid black line is the deviation from the oxygen isotopic equilibrium (between $HCO_3^-$ and $H_2O$) that is imprinted in the precipitated calcite at time t.

**Figure 4:** The temporal evolution of the ISOLUTION-modelled deviation of $\delta^{18}O_{calcite}$ from the equilibrium $\delta^{18}O^{eq}_{calcite}$ value with respect to $\delta^{18}O_{H2O}$ (dashed line) that results from calcite precipitation and the oxygen isotope exchange between $HCO_3^-$ and $H_2O$. $\delta^{18}O^{eq}_{calcite}$ is calculated using the Kim and O'Neil (1997) temperature relationship for $^{18}\alpha_{calcite/H2O}$ (Fig. 3). We note that the deviation of $\delta^{18}O_{calcite}$ from $\delta^{18}O^{eq}_{calcite}$ is independent of the used oxygen isotope fractionation factor. The temporal evolution of $\delta^{18}O_{calcite}$ is illustrated for three different $pCO_{2.karst}$ values (3000, 5000 and 10,000 ppm) while all other ISOLUTION parameters such as temperature (10 °C) and the $pCO_2$ value of the fracture or cave where calcite precipitation takes place (2000 ppm) were kept constant. $\delta^{18}O_{calcite}$ is calculated from the $\delta^{18}O$ values of $HCO_3^-$, thus retracing the temporal evolution of the solution's $HCO_3^-$ $\delta^{18}O$ value at a given point of time. The temporal evolution of the $\delta^{18}O_{calcite}$ values reveal that the maximum of the deviation can be up to c. 1.8 ‰ for the modelled scenarios and depend on the super-saturation of the percolating solution. Since the precipitation rate is faster than the oxygen isotope exchange between $HCO_3^-$ and $H_2O$, the increase of the oxygen isotope disequilibrium between the precipitated calcite and water is much faster than the oxygen isotope re-equilibration, which may take up to several days depending on temperature (Tab. 1).

**Figure 5.** Evolution of the oxygen isotope fractionation factor between water and calcite, $1000\ln(^{18}\alpha_{calcite/H2O})$, calculated from the inorganic calcite precipitation experiments illustrated in Fig. 2, adopted from Hansen et al. (2019). Note that the $\delta^{18}O$ value of $H_2O$ in these experiments is constant and not affected by calcite precipitation (Hansen et al., 2019). Illustrated are three cases in which (a) no prior calcite precipitation (PCP) occurs (Stalagmite 1), PCP occurs in the karst (Stalagmite 2) and (b) in the cave on stalactites (Stalagmite 3 in panel a). When no PCP occurs, the drip water is in oxygen isotope equilibrium. If PCP occurs, e.g. in a karst fracture or on the surface of a stalactite, the increase in calcite $\delta^{18}O$ values and the constant $H_2O$ $\delta^{18}O$ values provoke that calculated $1000\ln(^{18}\alpha_{calcite/H2O})$ values progressively increase in dependence on the residence time. For the illustrated experiments, the order of magnitude in changes in the $1000\ln(^{18}\alpha_{calcite/H2O})$ values is about the same as observed for the variability of published $1000\ln(^{18}\alpha_{calcite/H2O})$. The main difference between PCP in the karst and on the surface of a stalactite is the possibility for the oxygen isotope re-equilibration after PCP occurred in the karst. Depending on the residence time, the oxygen isotope disequilibrium can get balanced again by the continuous oxygen isotope exchange between $H_2O$ and $HCO_3^-$ (Fig. 4) and may reach oxygen isotope equilibrium conditions again. This is not the case if PCP occurs on the surface of stalactites, because secondary calcite is 'precipitated' on the surface of stalagmites or on watch classes after the drop falls from the stalactite tip.

## Table Caption

**Table 1:** The table lists the precipitation time constant ($\tau_P$) and the time constant for the oxygen isotope exchange between $H_2O$ and $HCO_3^-$ ($\tau_B$) for five different temperatures ranging from 0 °C and 20 °C. If the $HCO_3^-$ is not in equilibrium with $H_2O$, it requires approximately 4 times $\tau_B$ to re-establish the isotopic equilibrium of $HCO_3^-$ with respect to $H_2O$.

**Table 2:** Calculation of the degree of PCP on a stalactite surface in dependence on the drip interval (1, 120 and 1800 s) and the geometry of the stalactite at 10 °C. For this calculation it is assumed that the cave water flows down evenly as a fluid film on the entire stalactite surface and that the stalactite has a parabolic shape $f(x)=a\bullet x^2$, where $x$ is the radius of the stalactite that varies in the interval from 0 to $r_{max}$ ($r_{max}$ = 2.5 and 5 cm) and $a$ (a = 6 and 12) is a scaling parameter that links the radius and the length of the stalactite. $r_{max}$ is the maximum radius at the stalactites base, where it is connected to the cave ceiling. In detail, the degree of PCP is calculated using the exponential law ($1-e^{-\Delta t/\tau_{PCP}}$), where Δt is the residence time of the drip water on the stalactite surface and $\tau_{PCP}$ the PCP precipitation time constant. To calculate the residence time of a water parcel on the stalactite surface, the flow velocity of the water is calculated by $v=0.06\ cm/s\bullet (Q^2 sin(\varphi)/x^2)^{1/3}$ (Eq. (5) in Short et al., 2005), where $Q$ is the volumetric drip water flux in $cm^3/h$ (estimated by the drip interval and a drip volume of 0.1 ml), $\varphi$ is the tangent angle and $x$ the radius. Therefore, the residence time is the time interval that the water parcel needs to flow down from the base of the stalactite at the cave ceiling to the tip of the stalactite. The thickness $h$ of the drip water layer on the stalactite surface is given by $h=11\ \mu m \bullet (Q/x/sin(\varphi))^{1/3}$ (Eq. (4) in Short et al., 2005). $h_{min}$ and $h_{max}$ are the minimum and maximum thickness of the drip water layer on the stalactite surface, whereat $h_{median}$ and $h_{95}$ state the median and 95 % quantile of the drip water layer. The prior calcite precipitation time constant $\tau_{PCP}$ is given by $h/\alpha$, where

$\alpha$ is the temperature dependent precipitation rate of calcite at thickness $h$ of the fluid layer (e.g. Kaufmann, 2003). $\alpha$ is estimated for the median thickness of the fluid layer $h_{median}$, using the known $\alpha$ value at 10 °C for a thickness of the fluid layer of 100 µm (Baker et al., 1998) (the typical thickness of the fluid layer on stalagmite surfaces). Because $\alpha$ is decreasing with decreasing thickness of the fluid layer (Baker et al., 1998), $\alpha$ is rescaled by a factor of $h_{median}$/100 µm to account for the thinner fluid layer on stalactite surfaces. This is a very conservative estimate, because the precipitation rate decreases at smaller rates if the thickness of the fluid layer deceases (e.g. Fig. 2 in Baker et al., 1998).

| Temperature (°C) | $\tau_P$ (s) | $\tau_B$ (s) | 4 times $\tau_B$ (days) |
|---|---|---|---|
| 0 | 2064 | 125,700 | 5.8 |
| 5 | 1124 | 65,900 | 3.1 |
| 10 | 726 | 35,300 | 1.6 |
| 15 | 490 | 20,000 | 0.9 |
| 20 | 338 | 10,600 | 0.5 |

Table 1

Table 2

| $r_{max}$ (cm) | a | drip interval (s) | length of stalactite surface (cm) | residence time $\Delta t$ (s) | $h_{min}$ (μm) | $h_{max}$ (μm) | $h_{median}$ (μm) | $h_{95}$ (μm) | $\tau_{PCP}$ (s) | Degree of PCP (%) |
|---|---|---|---|---|---|---|---|---|---|---|
| 2.5 | 6 | 1 | 37.0 | 16.7 | 57.7 | 265.1 | 72.2 | 146.6 | 726 | 2.27 |
| 2.5 | 6 | 120 | 37.0 | 405.2 | 11.7 | 53.7 | 14.6 | 29.7 | 726 | 42.78 |
| 2.5 | 6 | 1800 | 37.0 | 2464.7 | 4.7 | 21.8 | 5.9 | 12.0 | 726 | 96.65 |
| 2.5 | 12 | 1 | 73.6 | 33.2 | 57.7 | 231.9 | 72.2 | 142.7 | 726 | 4.48 |
| 2.5 | 12 | 120 | 73.6 | 808.8 | 11.7 | 47.0 | 14.6 | 28.9 | 726 | 67.19 |
| 2.5 | 12 | 1800 | 73.6 | 4919.1 | 4.7 | 19.1 | 5.9 | 11.7 | 726 | 99.89 |
| 5 | 12 | 1 | 297.2 | 213.8 | 45.8 | 231.9 | 57.5 | 117.6 | 726 | 25.52 |
| 5 | 12 | 120 | 297.2 | 5202.3 | 9.3 | 47.0 | 11.7 | 23.9 | 726 | 99.92 |
| 5 | 12 | 1800 | 297.2 | 31,641.4 | 3.8 | 19.1 | 4.7 | 9.7 | 726 | 100 |

Figure 1

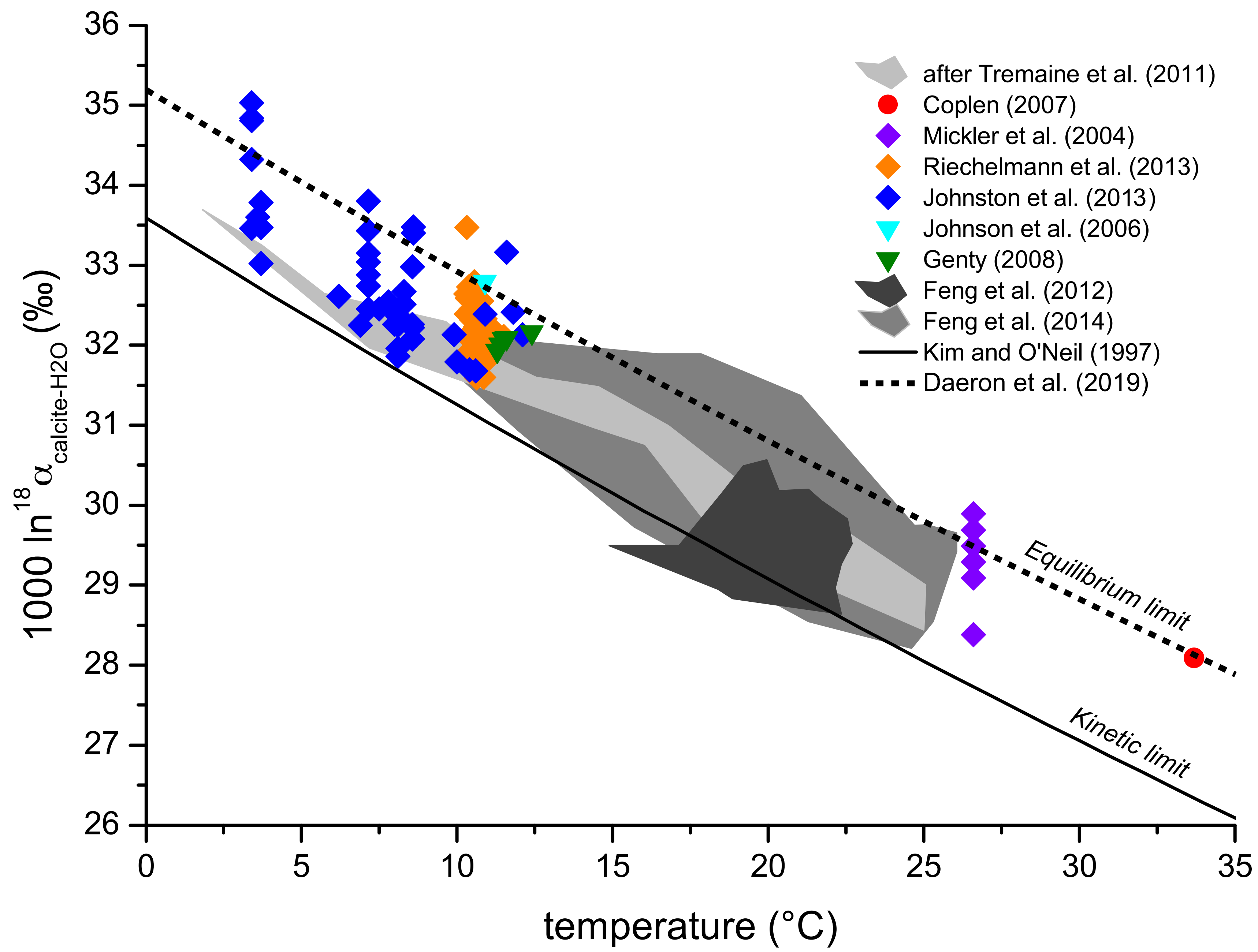

36
35
34
33
32
31
30
29
28
27
26
1000 ln$^{18}\alpha_{calcite-H2O}$ (‰)
0
5
10
15
20
25
30
35
temperature (°C)
after Tremaine et al. (2011)
Coplen (2007)
Mickler et al. (2004)
Riechelmann et al. (2013)
Johnston et al. (2013)
Johnson et al. (2006)
Genty (2008)
Feng et al. (2012)
Feng et al. (2014)
Kim and O'Neil (1997)
Daeron et al. (2019)
Equilibrium limit
Kinetic limit

Figure 2

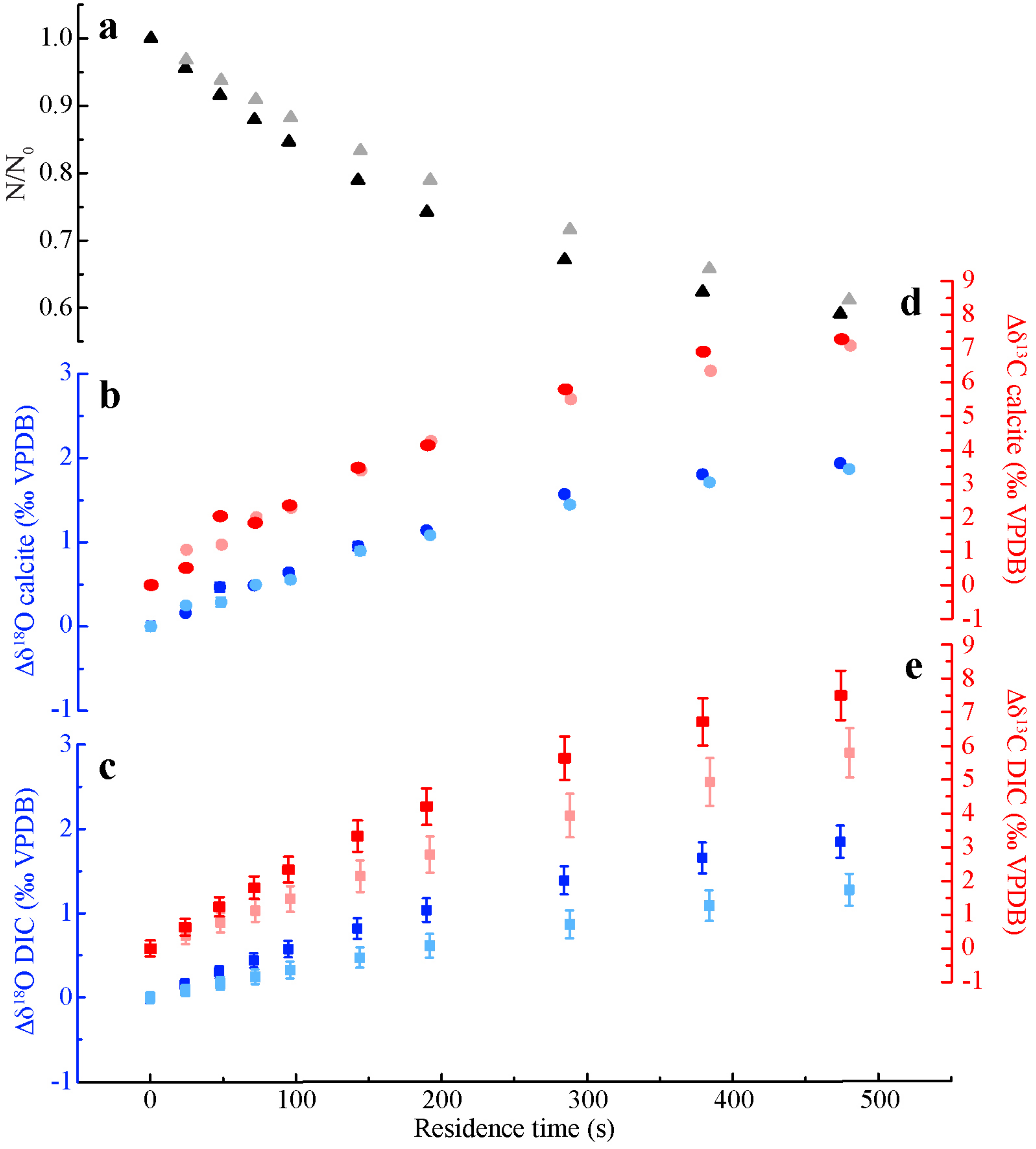

Figure 3

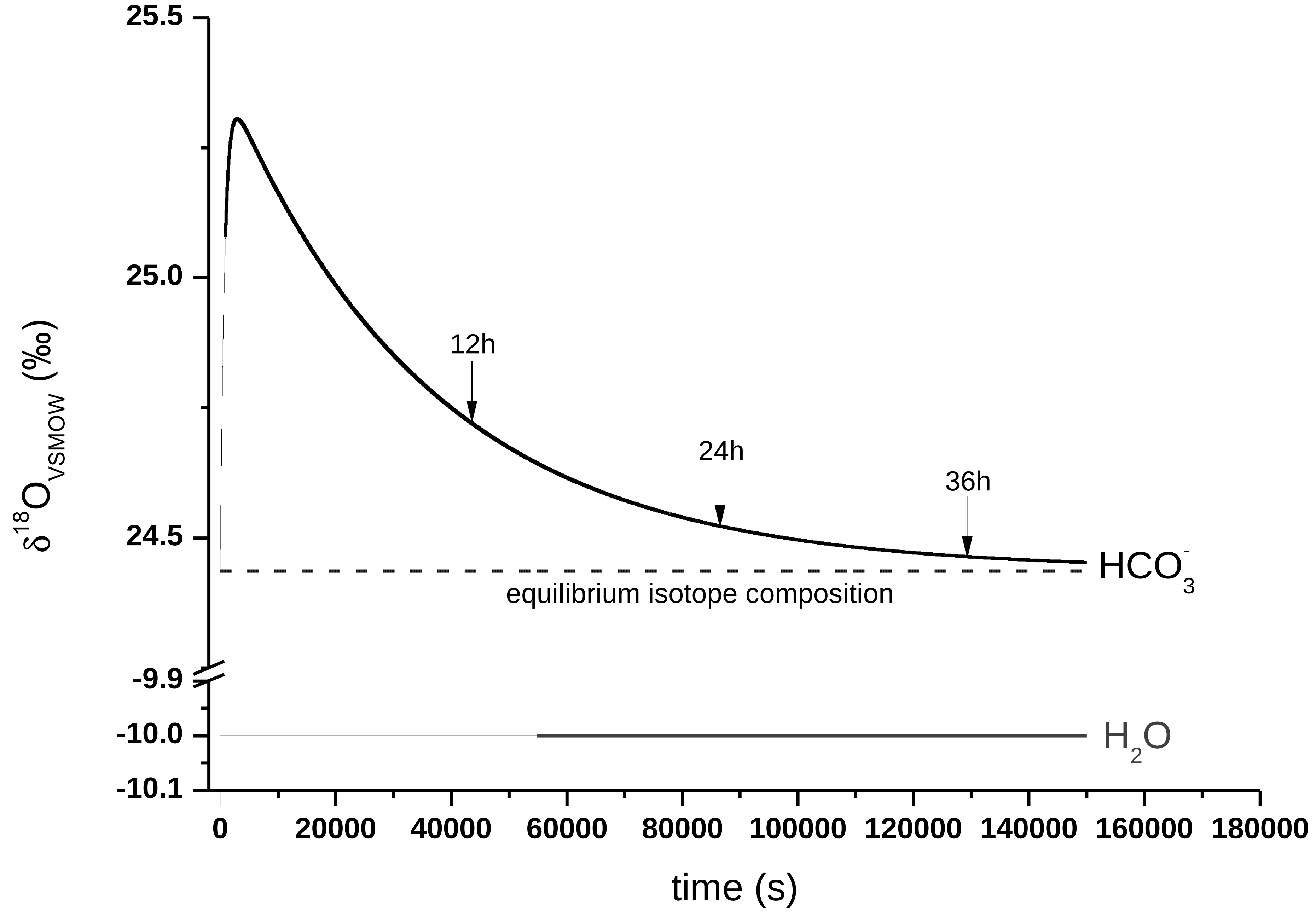

Figure 4

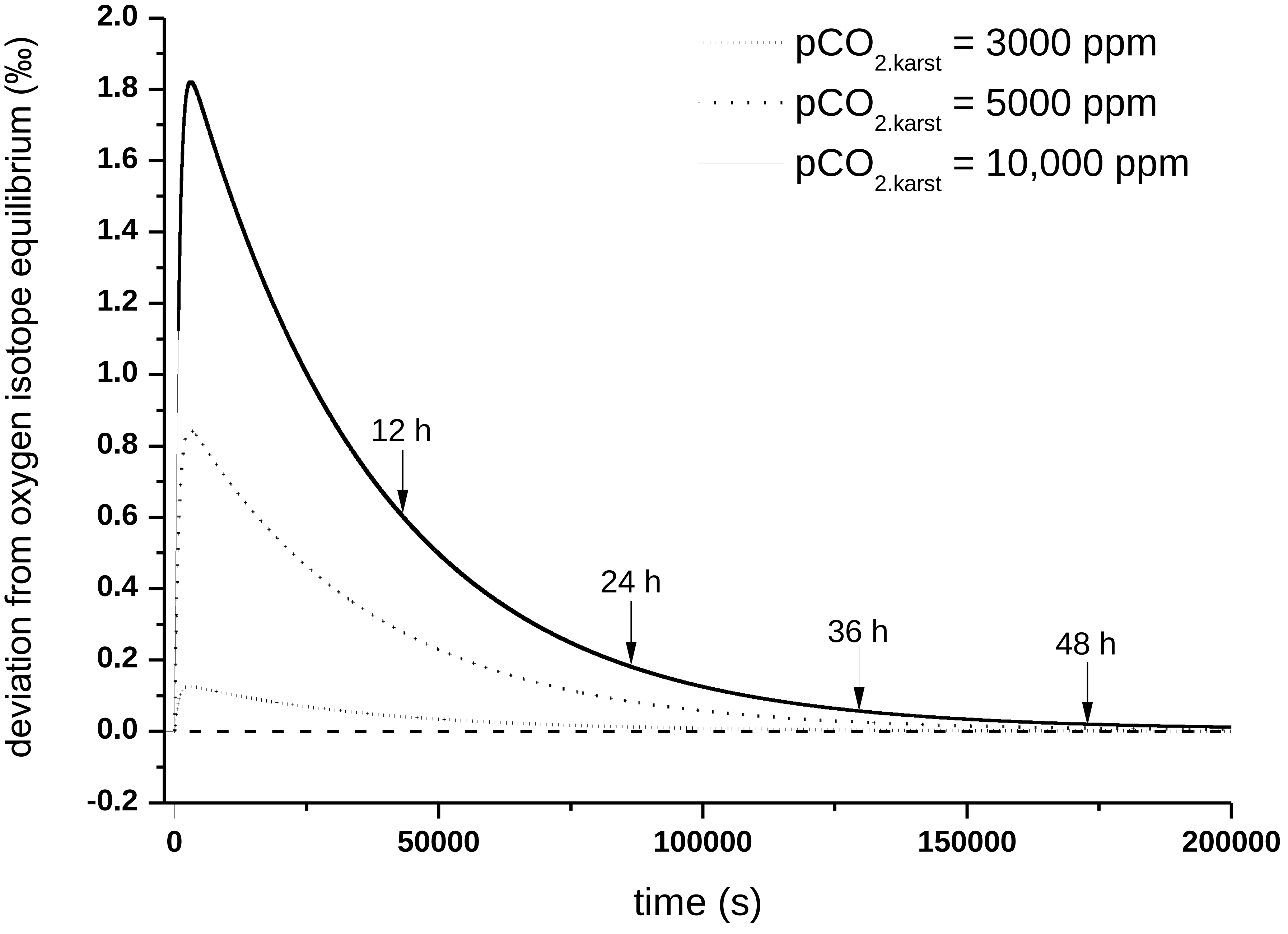

Figure 5

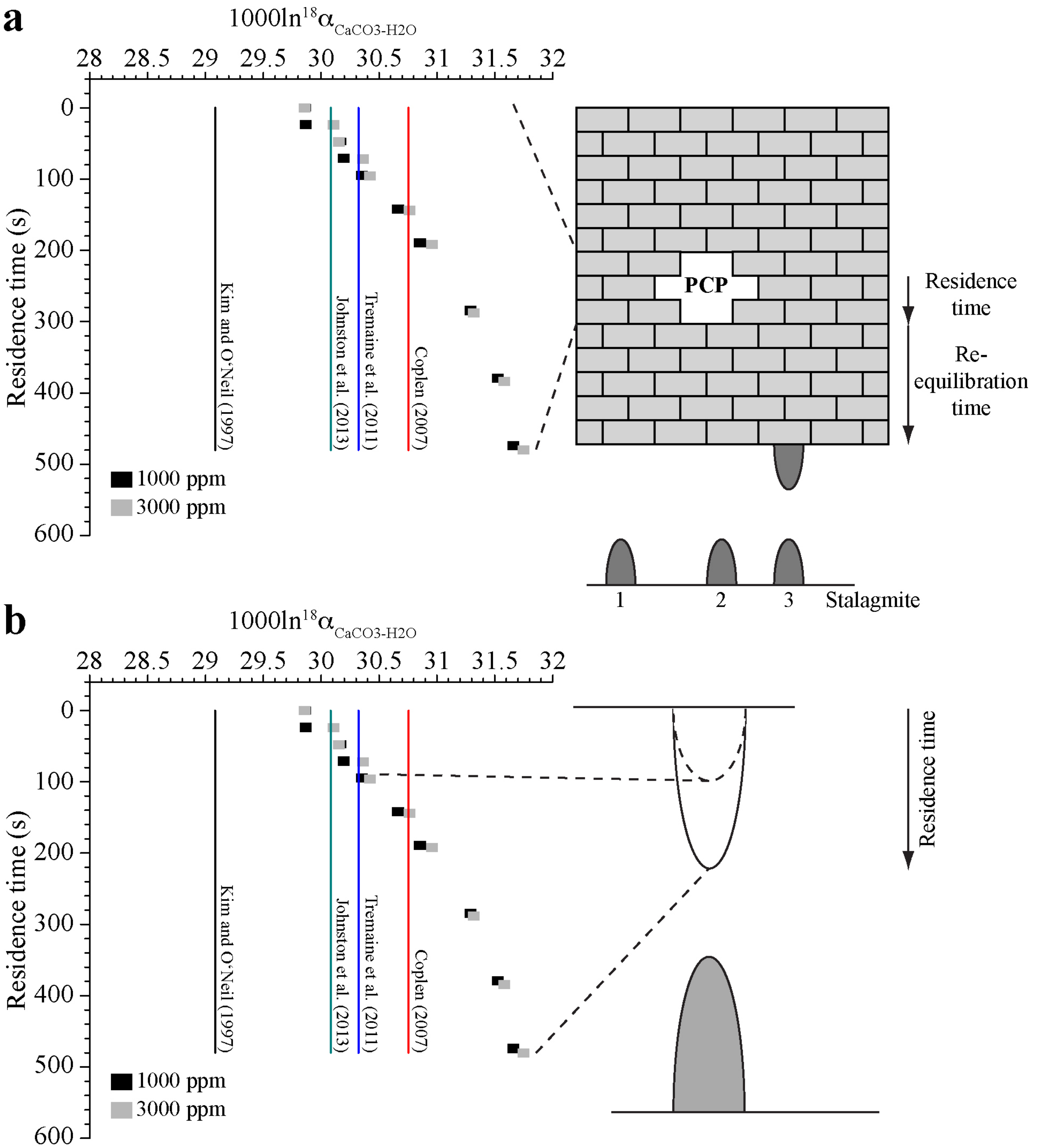
a
1000ln[18]α CaCO3-H2O
28 28.5 29 29.5 30 30.5 31 31.5 32
Residence time (s)
0 100 200 300 400 500 600
Kim and O'Neil (1997)
Johnston et al. (2013)
Tremaine et al. (2011)
Coplen (2007)
1000 ppm
3000 ppm
PCP
Residence time
Re-equilibration time
1 2 3 Stalagmite
b
1000ln[18]α CaCO3-H2O
28 28.5 29 29.5 30 30.5 31 31.5 32
Residence time (s)
0 100 200 300 400 500 600
Kim and O'Neil (1997)
Johnston et al. (2013)
Tremaine et al. (2011)
Coplen (2007)
1000 ppm
3000 ppm
Residence time